\documentclass[twocolumn,showpacs,preprintnumbers]{revtex4}
\usepackage{graphicx}
\usepackage{dcolumn}
\usepackage{bm}
\usepackage{subfigure}
\begin{document}
\newcommand{\ds}{\displaystyle}
\newcommand{\be}{\begin{equation}}
\newcommand{\en}{\end{equation}}
\newcommand{\bea}{\begin{eqnarray}}
\newcommand{\ena}{\end{eqnarray}}
\title{Friedmann-Robertson-Walker like cosmologies with spherical symmetry}
\author{Mauricio Cataldo}
\altaffiliation{mcataldo@ubiobio.cl} \affiliation{Departamento de
F\'\i sica, Facultad de Ciencias, Universidad del B\'\i o-B\'\i o,
Avenida Collao 1202, Casilla 5-C, Concepci\'on, Chile.}
\author{Fernanda Ar\'{o}stica}
\altaffiliation{ferarostica@udec.cl} \affiliation{Departamento de F\'{\i}sica, Universidad de Concepci\'{o}n,\\
Casilla 160-C, Concepci\'{o}n, Chile.}
\author{Sebastian Bahamonde}
\altaffiliation{sbahamonde@udec.cl} \affiliation{Departamento de F\'{\i}sica, Universidad de Concepci\'{o}n,\\
Casilla 160-C, Concepci\'{o}n, Chile.}\date{\today}
\begin{abstract}
We reconsider the cosmological model discussed by Sung-Won Kim
[Phys. Rev D {\bf 53}, 6889 (1996)] in the context of
Friedmann-Robertson-Walker cosmologies with a traversable wormhole,
where it is assumed that the matter content is divided into two
parts: the cosmic part that depends on time only and the wormhole
part that depends on space only. The cosmic part obeys the
barotropic equation of state $p_c=\gamma \rho_c$. The complete
analysis requires further care and reveals more interesting results
than what was previously shown by the author. They can be readily
applied to the evolution of a large class of cosmological models
which are more general than FRW models.

\vspace{0.5cm} 
\end{abstract}
\smallskip
\maketitle 
\section{Introduction}
 Wormhole models, and in particular evolving ones, have attracted
attention in the past few decades. In the Ref.~\cite{Kim} the author
considers a Friedmann-Robertson-Walker (FRW) cosmological model with
a traversable wormhole, by assuming that the matter is divided into
two parts: the cosmic part that depends on time only and the
wormhole part that depends on space only.

Specifically, the author considers the metric element of the
wormhole in a FRW universe in the form
\begin{eqnarray}\label{evolving wormhole915}
ds^2=-dt^2+  R(t)^2 \left( \frac{dr^2}{1-kr^2-\frac{b(r)}{r}}+r^2 d
\Omega^2 \right),
\end{eqnarray}
where $d \Omega^2=d\theta^2+sin^2 \theta d \phi^2$, $R(t)$ is the
scale factor of the universe and $k$ is the sign of the curvature of
spacetime: $+1,0,-1$. The Einstein equation are written in the form
\begin{eqnarray}\label{00}
\kappa \rho(t,r)=3 H^2+\frac{3k}{R^2}+\frac{b^{\prime}}{R^2 r^2}-\Lambda,  \\
\label{rr} \kappa \tau(t,r)=\left( 2\frac{\ddot{R}}{R}+ H^2
+\frac{k}{R^2} \right)+ \frac{b}{R^2
r^3} - \Lambda, \\
\label{11} \kappa P(t,r)= -\left(2 \frac{\ddot{R}}{R}+ H^2
+\frac{k}{R^2} \right) + \frac{b-r b^{\prime}} {2 R^2 r^3}+\Lambda,
\label{22}
\end{eqnarray}
where $\kappa=8 \pi G$, $H=\dot{R}/R$, and a prime and an overdot
denote differentiation with respect to $r$ and $t$ respectively. We
have added the cosmological constant $\Lambda$ to the Einstein
equations. Here $\rho(t,r)$, $\tau(t,r)$, $P(t,r)$ are the mass
energy density, radial tension and lateral pressure. Note that for
the tension we have that $\tau=-P^r$, where $P^r(t,r)$ is the radial
pressure.

Since the the mass energy density, radial and lateral pressures
depend on both $t$ and $r$, the following ansatz for matter parts is
introduced by the author:
\begin{eqnarray}\label{uno}
R^2(t) \rho(t,r)=R^2(t) \rho_c(t)+\rho_w(r),  \\
R^2(t) P^r(t,r)=R^2(t) P_c(t)+P^r_w(r),  \label{dos} \\
R^2(t) P(t,r)=R^2(t) P_c(t)+P_w(r). \label{tres}
\end{eqnarray}
The subscripts $c$ and $w$ indicate the cosmological and wormhole
parts respectively. Note that, without any loss of generality, the
Eq.~(\ref{dos}) has been rewritten by using the radial pressure
$P^r(t,r)$ instead of the radial tension $\tau(t,r)$, as made in
Ref.~\cite{Kim}. The cosmological part is represented by an
isotropic pressure $P_c$.

The Eqs.~(\ref{uno})-(\ref{tres}) allow us to separate the Einstein
equations~(\ref{00})-(\ref{22}) into two parts
\begin{eqnarray}\label{S00}
R^2 \left(\kappa \rho_c-\left[3 H^2+\frac{3k}{R^2}-\Lambda
\right]\right)=\frac{b^{\prime}}{r^2}-\kappa
\rho_w=l,  \\
R^2 \left(\kappa P_c+ 2\frac{\ddot{R}}{R}+ H^2
+\frac{k}{R^2}-\Lambda
\right)=- \frac{b}{r^3} - \kappa P^r_w=m, \label{S11} \\
R^2 \left(\kappa P_c+ 2\frac{\ddot{R}}{R}+ H^2
+\frac{k}{R^2}-\Lambda \right)=\frac{b-r b^{\prime}}{2 r^3} - \kappa
P_w=m, \nonumber \\  \label{S22}
\end{eqnarray}
where $l$ and $m$ are arbitrary separation constants. The separation
constants of Eqs.~(\ref{S11}) and~(\ref{S22}) are equal due to the
form of the cosmological part. Notice that for the cosmological part
from Eqs.~(\ref{S00}) and~(\ref{S11}) we can construct the equation
\begin{eqnarray}\label{generalized CE}
\dot{\rho}_c+3H(\rho_c+P_c)=\frac{l+3m}{\kappa}\dot{R}R^{-3}.
\end{eqnarray}
This is a generalized conservation equation for the cosmological
part.

{\bf Case $l=-3m$}: For parameter values satisfying this constraint
we obtain the standard conservation equation for FRW metric. In this
case Eqs.~(\ref{S00})-(\ref{S22}) allow us to write the cosmological
part in the form
\begin{eqnarray}\label{CS00}
3 H^2+\frac{3(k-m)}{R^2} =\kappa \rho_c+\Lambda,  \\
- 2\frac{\ddot{R}}{R}- H^2 -\frac{k-m}{R^2} =\kappa P_c-\Lambda,
\label{CS11}
\end{eqnarray}
while the wormhole part is given by
\begin{eqnarray}\label{WS00}
\frac{b^{\prime}}{r^2}-\kappa
\rho_w=-3m,  \\
- \frac{b}{r^3} - \kappa P^r_w=m, \label{WS11} \\
\frac{b-r b^{\prime}}{2 r^3} - \kappa P_w=m. \label{WS22}
\end{eqnarray}
From Eqs.~(\ref{WS00})-(\ref{WS22}) we obtain that
$\rho_w+P^r_w+2P_w=0$. This relation implies that if we require that
one of the pressures has a barotropic equation of state then the
another one also has the same type of equation of state. So, for
example, if we require that the radial pressure has the form
$P^r_w=\omega_r \rho_w$, then the lateral pressure is given by the
barotropic equation of state $P_w=-(1+\omega_r) \rho_w/2$.

From the above Eqs.~(\ref{CS00})-(\ref{WS22}) it is clear that the
isotropic cosmological matter distribution $\rho_c$ determines the
behavior of the scale factor $R(t)$, while the relevant metric
function $b(r)$ is determined by the matter distribution $\rho_w$.
Notice that Eqs.~(\ref{CS00}) and~(\ref{CS11}) are a generalization
of the FRW equations, from which if $m=0$ we obtain the standard FRW
equations. In this case is still valid the relation
$\rho_w+P^r_w+2P_w=0$.

Notice that in Eqs.~(\ref{WS00})-(\ref{WS22}) the energy density,
the radial and lateral pressures are written through the relevant
metric function $b(r)$. However, equations of the wormhole part may
be rewritten through energy density or one of the pressures. We
shall write these equations through the radial pressure $P_w^r(r)$.
Thus Eqs.~(\ref{WS00})-(\ref{WS22}) may be rewritten in the
following differential form:
\begin{eqnarray}\label{bder}
b(r)=-mr^3-\kappa r^3 P^r_w, \\ \label{rhoder} \rho_w(r)=-r^{-2}
\left(r^3 P^r_w  \right)^\prime, \\ \label{pder}P_w(r)=
P^r_w+\frac{r}{2} {P_w^r}^{\prime}.
\end{eqnarray}
This allows us to write the metric~(\ref{evolving wormhole915}) as
\begin{eqnarray}\label{evolving wormhole9150}
ds^2=-dt^2+  R(t)^2  \times \nonumber \\ \left(
\frac{dr^2}{1-(k-m)r^2+\kappa r^2 P_w^r(r)}+r^2 d \Omega^2 \right).
\end{eqnarray}
This analysis reveals that a large class of cosmological models,
given by the metric~(\ref{evolving wormhole9150}), are determined by
the arbitrary function $P_w^r(r)$ and the matter distribution
$\rho_c(t)$ which, with the help of the Friedmann-like
Eqs.~(\ref{CS00}) and~(\ref{CS11}), determines the scale factor
$R(t)$.

As a simple example let us consider the barotropic case
$P_w^r=\omega_r \rho_w$. Thus from Eqs.~(\ref{rhoder})
and~(\ref{pder}) we have that $\rho_w=C r^{-3-1/\omega_r}$ and
$2P_w=-C(\omega_r+1) \, r^{-3-1/\omega_r} $ respectively. This
specific class of solutions, for $m=0$, was studied in
Ref~\cite{Cataldo}. In this case we can consider any isotropic
matter distribution $\rho_c$ in order to define the scale factor
$R(t)$.


{\bf Case $l \neq -3m$}: For this parameter relation we can consider
that Eq.~(\ref{generalized CE}) is the conservation equation of a
viscous fluid, so the cosmological part will describe viscous
cosmologies. In order to construct this interpretation let us make
$l=3 \tilde{l}$. Thus, the cosmological part may be written in the
form
\begin{eqnarray}\label{VCS00}
3 H^2+\frac{3(k+\tilde{l})}{R^2}=\kappa \rho_c+\Lambda,  \\
\kappa P_c+ 2\frac{\ddot{R}}{R}+ H^2
+\frac{k+\tilde{l}}{R^2}-\Lambda=\frac{m+\tilde{l}}{R^2},
\label{VCS11}
\end{eqnarray}
while the wormhole part is represented in the form
\begin{eqnarray}\label{VWS00}
\frac{b^{\prime}}{r^2}-\kappa
\rho_w=3\tilde{l},  \\
- \frac{b}{r^3} - \kappa P^r_w=m, \label{VWS11} \\
\frac{b-r b^{\prime}}{2 r^3} - \kappa P_w=m. \label{VWS22}
\end{eqnarray}
From Eqs.~(\ref{VWS00})-(\ref{VWS22}) we have that $\kappa
\rho_w+\kappa P^r_w+2 \kappa P_w=-3(\tilde{l}+m)$. If we require a
barotropic radial pressure $P^r_w=\omega_r \rho_w$, the lateral
pressure has a generalized equation of state of the form $\kappa
P_w=-3(\tilde{l}+m)/2-(1+\omega_r) \kappa \rho_w/2$.

In order to consider the viscous models we may introduce the
cosmological viscous pressure in the form $P_{visc}=P_c-3 H \xi$,
where $\xi$ is the bulk viscosity. Thus from Eq.~(\ref{generalized
CE}) we obtain that
\begin{eqnarray}
\xi(t)=\frac{\tilde{l}+m}{3 \kappa R \dot{R}}.
\end{eqnarray}
It is interesting to remark that for $\tilde{l}=0$ we obtain
standard open, flat and closed viscous FRW cosmologies.

In this case, from Eqs.~(\ref{VWS00})-(\ref{VWS22}), we can write
the relevant metric function $b(r)$, the radial and lateral
pressures trough the energy density in the following integral form:
\begin{eqnarray} \label{int1}
b(r)=C_{1}+\tilde{l}r^{3}+\kappa\int{r^{2}\rho_{\omega}(r)dr},\\
\kappa
P^{r}_{\omega}(r)=-\frac{C_1}{r^{3}}-(\tilde{l}+m)-\frac{\kappa}{r^{3}}\int{\rho_{\omega}(r)r^{2}dr}
\label{int1a} \\ \kappa
P_{\omega}(r)=\frac{C_1}{2r^{3}}-(\tilde{l}+m)-\frac{\kappa}{2}\rho_{\omega}(r)+
\nonumber \\ \frac{\kappa}{2r^{3}}\int{r^{2} \rho_{\omega}(r)dr}.
\label{int1b}
\end{eqnarray}
These equations for the constraint $\tilde{l}=-m$ may be directly
rewitten in the form of Eqs.~(\ref{bder})-(\ref{pder}).

This allows us to write the metric~(\ref{evolving wormhole915}) in
the following form:
\begin{eqnarray}\label{evolving wormhole9150V}
ds^2=-dt^2+  R(t)^2  \times \nonumber \\ \left(
\frac{dr^2}{1+\frac{C_1}{r}-(k+\tilde{l})r^2+\frac{\kappa}{r} \int
r^2 \rho_w(r) dr}+r^2 d \Omega^2 \right).
\end{eqnarray}
In this case the analysis reveals that a large class of viscous
cosmological models, given by the metric~(\ref{evolving
wormhole9150V}), are determined by the arbitrary function
$\rho_w(r)$ and the isotropic matter distribution $\rho_c(t)$ which,
with the help of the Friedmann-like Eqs.~(\ref{VCS00})
and~(\ref{VCS11}), determines the scale factor $R(t)$.

As a simple application let us consider the cosmological model
studied in Ref.~\cite{Kim}, defined by the equation of state
$p_c=\gamma \rho_c$ for the cosmological part. In this case the
conservation equation~(\ref{generalized CE}) (with $l=3 \tilde{l}$)
gives for the cosmic energy density
\begin{eqnarray}
\kappa \rho_c&=& C
R(t)^{-3(\gamma+1)}+3\frac{\tilde{l}+m}{1+3\gamma}R^{-2}(t),
\end{eqnarray}
where $C$ is an integration constant. By putting this relationship
into Eq.~(\ref{VCS00}) we find that the scale factor satisfies the
differential equation
\begin{eqnarray}\label{915}
3 H^2+\frac{3  \alpha}{R^2}=C R(t)^{-3(\gamma+1)}+\Lambda,
\end{eqnarray}
where $\alpha=k+(3 \gamma \tilde{l}-m)/(1+3 \gamma)$. The particular
solution for Eq.~(\ref{915}) considered in Ref.~\cite{Kim} is
expressed by the scale factor
\begin{eqnarray}\label{SF15}
R(t)=R_0 t^{2/(3(\gamma+1))},
\end{eqnarray}
for $\gamma \neq -1$. This implies that the constraint
$\alpha=\Lambda=0$ must be fulfilled. The bulk viscosity and energy
density take the form
\begin{eqnarray}
\kappa \xi=\frac{(\tilde{l}+m)(1+\gamma)}{2 R^2_0} \, t^{\frac{3
\gamma-1}{3(\gamma+1)}}, \\
\kappa \rho_c= \rho_{c0} t^{-2} + 3 \frac{\tilde{l}+m}{(1+3 \gamma)
R_0^{2}} \, t^{-\frac{4}{3(1+\gamma)}},
\end{eqnarray}
respectively, where $\rho_{c0}=4/(3 \kappa (1+\gamma)^2)$. Note that
a careful analysis indicates that the scale factor~(\ref{SF15})
defines the expansion for a large class of spacetimes expressed by
the metric~(\ref{evolving wormhole9150V}). Independently, the
spatial part of this metric is defined exclusively by the matter
content $\rho_w(r)$. Let us consider a particular form of the
spatial part. If we require, for example, the condition $P^r_w=\beta
\rho_w$, from Eq.~(\ref{int1a}) we find that
\begin{eqnarray}\label{rrrr}
\kappa \rho_w=C r^{-\frac{3 \beta+1}{\beta}}-3 \frac{\tilde{l}+m}{3
\beta+1},
\end{eqnarray}
where $C$ is a constant of integration. By putting this expression
back into Eq.~(\ref{int1a}) we find that $C_1=0$ in order to have
the required barotropic equation of state $P^r_w=\beta \rho_w$. From
Eqs.~(\ref{int1b}) and~(\ref{rrrr}) we obtain for lateral pressure
\begin{eqnarray}\label{ppprrr}
\kappa P_w=-\frac{C (\beta+1)}{2} r^{-\frac{3
\beta+1}{\beta}}-3\frac{\beta (\tilde{l}+m)}{3 \beta+1}.
\end{eqnarray}
In this case the metric takes the following form:
\begin{eqnarray}
ds^2=-dt^2+  R^2_0 t^{\frac{4}{3(\gamma+1)}}  \times \nonumber \\
\left( \frac{dr^2}{1-\tilde{\alpha} r^2-C \beta
r^{-\frac{1+\beta}{\beta}}}+r^2 d \Omega^2 \right),
\end{eqnarray}
where $\tilde{\alpha}=-(l+k)-{\frac {l+m}{3\,\beta+1}}$. From
Eq.~(\ref{ppprrr}) we conclude that the lateral pressure has a
barotropic equation of state only for a vanishing bulk viscosity,
since $\tilde{l}=-m$ must be required. The specific Kim's solution
is obtained by putting $\beta_{our}= -(1+2 \beta_{Kim})$ and
$\tilde{l}=-m$ into Eqs.~(\ref{rrrr}) and~(\ref{ppprrr}).

As another example, we can consider a direct generalization of the
previously discussed viscous model by including the constant $C_1$.
As we can see from Eqs.~(\ref{int1})-(\ref{int1b}), in this case the
radial and lateral pressures do not obey barotropic equations of
state with constant state parameters.

In conclusion, the gravitational configuration defined by
Eqs.~(\ref{evolving wormhole915})-(\ref{22}) may be separated, with
the help of the ansatz~(\ref{uno})-(\ref{tres}), into a two
independent gravitational systems: one with Einstein field equations
describing a static gravitational field and another describing a
time-dependent gravitational field. In general, the static field
part corresponds to a large class of spherically symmetric solutions
of the Einstein equations, which may represent zero-tidal-force
wormholes or naked singularities, among others static
configurations, while the time-dependent field part represents
Friedmann-like field equations for FRW models with any curvature k.
Notice that they do not significantly differ from the standard FRW
ones. For parameter values $l=m=0$, the Friedmann-like equations
become the standard FRW ones. These equations define the expansion
factor $R(t)$ of a large class of metrics~(\ref{evolving
wormhole915}), where the spatial part depends on the arbitrary
function $b(r)$. The presence of the matter content $\rho_w(r)$ is
essential to this spatial part, since from
Eqs.~(\ref{WS00})-(\ref{WS22}) we have for $\rho_w=P^r_w=P_w=0$ that
the arbitrary function $b(r)$ is trivially given by $b(r)=-m r^3$.
Thus, the standard FRW field equations determine the expansion not
only of an isotropic and homogeneous FRW metric, but also of more
general metrics given by Eq.~(\ref{evolving wormhole915}).

\section{Acknowledgements}
This work was supported by CONICYT through Grant FONDECYT N$^0$
1080530 and by the Direcci\'on de Investigaci\'on de la Universidad
del Bio-B\'\i o through grants N$^0$ DIUBB 121007 2/R and N$^0$
GI121407/VBC (MC).

\end{document}